\documentclass [%
superscriptaddress,
preprint,
 amsmath,amssymb,
 showkeys
]{revtex4-1}

\usepackage{graphicx}
\usepackage{dcolumn}
\usepackage{bm}
\usepackage{color}
\usepackage{float}
\usepackage[utf8]{inputenc}

\usepackage{multirow}
\usepackage{hyperref}
\usepackage{amsmath}
\usepackage{xr}

\usepackage{soul,xcolor}
\soulregister\cite7
\soulregister\ref7
\setstcolor{blue}
\begin{document}

\title{Spin-orbit enhanced carrier lifetimes in noncentrosymmetric semiconductors}

\author{Liang Z. Tan}
\affiliation{Department of Chemistry, University of Pennsylvania, Philadelphia, Pennsylvania 19104, USA}
\author{Andrew M. Rappe}
\affiliation{Department of Chemistry, University of Pennsylvania, Philadelphia, Pennsylvania 19104, USA}

\date{\today}

\begin{abstract}
We show that the carrier recombination rate of noncentrosymmetric materials can be strongly modified by spin-orbit coupling. Our proposed mechanism involves the separation of conduction and valence bands into their respective spin components, which changes the transition dipole moments between them. The change in the carrier recombination can be either positive or negative in sign, or vary depending on the location of carriers in the Brillouin zone. We have performed a large scale DFT screening study to identify candidate materials that display this effect. We have selected three materials, Pb$_4$SeBr$_6$, ReTe$_3$Br$_5$, and CsCu(BiS$_2$)$_2$, which span the range of behaviors, and discuss their electronic band structure in greater detail. We find transition dipole moment enhancement factors of up to three orders of magnitude, reflecting the physical impact of spin-orbit coupling on the carrier lifetime. Therefore, further explorations of the spin-orbit coupling and lattice symmetry could prove to be useful for manipulating the photophysics of materials.
\end{abstract}

\maketitle

\section{Introduction}

Radiative recombination of carriers is a major loss mechanism in many semiconductors. Long carrier lifetimes are desirable for collection of carriers in charge transport devices and for the observation of excited-state physics in general. In most solid-state materials, the radiative recombination rate~\cite{vanRoosbroeck54p1558,Wurfel82p3967,Trupke03p4930} is of the order of 10$^{-15}$--10$^{-13}$ cm$^3$s$^{-1}$, and it is mostly regarded as an unavoidable loss mechanism.    

In this paper, we show that the carrier recombination rate is strongly affected by the presence of spin-orbit coupling, and that it can be quenched by choosing spin-orbit materials where the spin degree of freedom can imposes selection rules on the lowest-energy interband transitions. We present a mechanism by which spin-orbit coupling either enhances or reduces the carrier lifetime, by separating the energy bands into different spin components, and hence modifying the radiative recombination rate. 

This mechanism was first proposed in the halide perovskites~\cite{Zheng15p7794,Etienne16p1638} as an explanation for the anomalously long carrier lifetimes in these materials~\cite{Stranks13p341,Xing13p344,Ponseca14p5189}. The halide perovskites, which have recently emerged as efficient solar cell materials~\cite{NREL15,Fan15p18809,Zheng15p4862,deAngelis17p922}, contain large spin-orbit coupling due to the heavy Pb and I atoms and are thought to display dynamically and locally  broken inversion symmetry~\cite{Poglitsch87p6373,Leguy15p7124,Ma14p248,Yaffe17p136001}. This results in a Rashba-type band structure~\cite{Rashba88p175}, which has been theoretically proposed in halide perovskites~\cite{Even13p2999,Kim14p6900,Kepenekian15p11557} and observed there in different types of experiments~\cite{Niesner16p126401,Yu16p3078,Isarov17pASAP}. 

We show that this mechanism of spin-orbit induced carrier lifetime enhancement is not limited to hybrid perovskites. We present materials that display this effect in their static ground-state structures. We show that, depending on material, the relative orientation of spins in the conduction and valence bands may be such that carrier recombination is suppressed, enhanced, or displays a more complex $k$-dependent behavior. Below, we provide examples of materials displaying each of these behaviors.

\section{Description of lifetime enhancement mechanism}

We first outline the mechanism behind spin-orbit induced carrier lifetime enhancement. This mechanism relies on electron and hole carriers occupying states where radiative recombination is dipole-forbidden (dark states). In this paper, we focus on the situation where the transition is dipole-forbidden in the spin sector, that is, because the electron and hole carriers have opposite (or nearly opposite) spin orientations. This situation is brought about by having a material with strong spin-orbit coupling, which causes a lifting of spin-degeneracies of bands. Electron and hole carriers in the band extrema would then be in particular spin states, with their spin directions being material-dependent quantities. For instance, the bands of Rashba-type materials~\cite{Rashba88p175} have a vortex-like arrangement of spins near high-symmetry points in the Brillouin zone. If the spins of electron and hole are anti-aligned (aligned), this would cause an enhancement (reduction) of carrier lifetime. In most materials, the electron and hole spins would not be perfectly aligned or anti-aligned; the degree of enhancement or reduction in carrier lifetime depends on their relative directions.

This effect requires a population of electron and hole carriers in dipole-forbidden states. Because of this selection rule, these carriers cannot be directly optically excited into these states, but can only arrive there by excitation into higher-energy states followed by relaxation into the dipole-forbidden states. It is necessary that these excitation and relaxation processes are not forbidden by selection rules and that they occur fairly rapidly, ensuring that a significant portion of carriers are in the dipole-forbidden states. 

\begin{figure}
\includegraphics[width=0.65\textwidth]{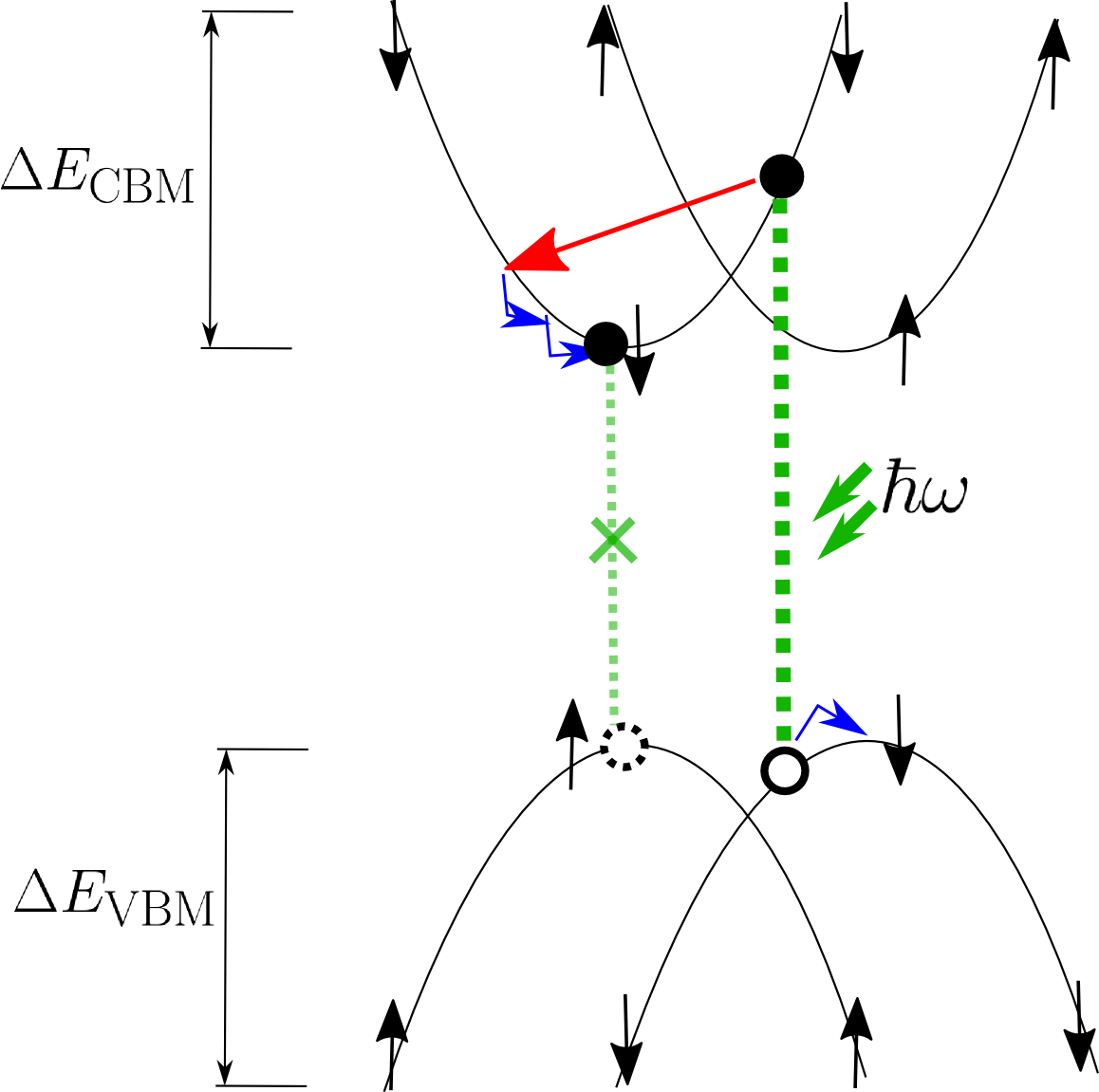}
  \caption{Schematic showing carrier excitation and relaxation
processes in a Rashba band structure. The black arrows indicate
the directions of the spins. After absorption of  photons (green, right), spin-aligned electrons and holes are created. 
The excited electrons relax to the
conduction band minimum due interband (red) and intraband (blue) phonon scattering processes, which usually do not flip the electron spin.
Similarly, the holes relax to the valence band
maximum without flipping their spins. After this thermalization process, radiative (vertical) recombination of electrons and holes is spin-forbidden due to opposite spin directions (crossed green line, left). The conduction band energy splitting ($\Delta E_{\text{CBM}}$) and valene band energy splitting ($\Delta E_{\text{VBM}}$) are defined in the figure. }\label{fig:recomb}
\end{figure}

In Fig.~\ref{fig:recomb}, we provide a concrete example of how this process occurs in a typical Rashba band structure~\cite{Rashba88p175}. This band structure, which occurs in noncentrosymmetric materials with strong spin-orbit coupling, contains conduction and valence bands which are each spin-split away from a high-symmetry point and degenerate at that point. The conduction band minimum (CBM) and valence band maximum (VBM) are therefore located away from the high-symmetry point. Upon excitation by a photon, an electron and a hole are created in bands with the same spin orientation (thick green arrow in Fig.~\ref{fig:recomb}). Shortly afterward, the electron and hole each relax via phonon scattering events, which occur primarily to states of the same spin orientation. We note that there are intraband (blue arrows in Fig.~\ref{fig:recomb}) as well as interband transitions (red arrow in Fig.~\ref{fig:recomb}) of this kind, allowing the carriers to relax to the CBM and VBM. A quasi-static distribution of carriers is thus established at the CBM and VBM. Here, their recombination is suppressed because of the opposing spin orientations of the CBM and VBM.    

This mechanism bears some similarity to dark excitons, which are bound electron-hole pairs with parallel spins, and  as a result have a long radiative lifetime~\cite{Mcfarlane09p093113}. However, there are several differences. Firstly, this mechanism pertains to free carriers instead of bound excitons, and is of greater relevance in situations where collection of current is important. Secondly, the dark exciton states are usually populated via an intersystem crossing~\cite{Webb70p4227}, which is mediated by the spin-orbit coupling. Here, the population of dark states is achieved by phonon scattering, which is typically a faster process~\cite{Bernardi14p257402,Tanimura16p161203} than intersystem crossings. Such phonon scattering processes are available only to free carriers. In both mechanisms, the size of the energy splitting opened by spin-orbit coupling determines the ratio of majority-spin to minority-spin states at a given temperature.

To summarize, the requirements for spin-orbit enhanced carrier lifetime are a strong spin-orbit interaction to ensure a large enough spin-splitting, a breaking of inversion symmetry to correctly order the spin-split bands to ensure that the CBM and VBM states are dipole forbidden, and fast carrier relaxation.

\section{methodology}

We screened the \textsc{Materials Project} database~\cite{Jain13p011002} for materials exhibiting a range of spin-orbit influences on carrier lifetime. We constrained our search to noncentrosymmetric materials with heavy elements (Z$>49$) which have non-magnetic ground states. Furthermore, we have restricted our search to thermodynamically stable materials, or metastable materials with decomposition energies of less than 0.1 eV/atom. We have chosen materials with fewer than 30 atoms per unit cell, for computational efficiency. This screening resulted in 421 candidate materials, for which we performed density functional theory (DFT) calculations of the band structure and transition dipole moments (TDMs). These calculations were done with the PBE density functional~\cite{Perdew96p3865}, using norm-conserving RRKJ pseudopotentials~\cite{Rappe90p1227}, and using a planewave basis set with kinetic energy cutoff of 60 Ry. Spin-orbit coupling was included at the fully-relativistic level for all calculations. A Monkhorst-Pack~\cite{Monkhorst76p5188} 8$\times$8$\times$8 $k$-point mesh was used for the self-consistent evaluation of the charge densities. 

From these DFT calculations, we have selected 16 materials with large spin-orbit splittings of the valence or conduction band (Table~\ref{table:1}). These materials have spin-orbit splittings larger than the room-temperature energy scale, and they therefore can host spin-resolved carrier populations. We have tabulated the relative ordering of spin-split conduction (CB1, CB2) and valence (VB1,VB2) bands. This was done by considering the TDMs of the four transitions from CB1 and CB2 to VB1 and VB2. If the lowest-energy transition (VB1$\rightarrow$CB1) has a lower TDM than the higher-energy transitions VB1$\rightarrow$CB2 and VB2$\rightarrow$CB1, carrier recombination will be suppressed. The highest-energy transition in this manifold of states, VB2$\rightarrow$CB2, often has a similar TDM to VB1$\rightarrow$CB1 because they have similar relative spin orientations (See Fig.~\ref{fig:recomb}). However, the highest transition does not affect the carrier population strongly if the spin-orbit splitting is larger than room temperature. Among the materials in Table~\ref{table:1}, we find examples where the ordering of bands either suppresses or enhances carrier lifetime, as well as more complex cases where the band structure contains both of these effects simultaneously, in different parts of the Brillouin zone. In the following, we will discuss representative examples of each case in more detail. 

\begin{table}[h!]
\centering
\begin{tabular}{|c | c c c c c|} 
 \hline
  & Direct gap (eV) & Indirect gap (eV) & $\Delta E_{\text{CBM}}$ (meV) & $\Delta E_{\text{VBM}}$(meV) & $\Delta\tau$ \\ 
 \hline\hline
ReTe$_3$Br$_5$ & 1.2157 & 1.1986 & 27.6 & 6.3 & +  \\
Pb$_4$SeBr$_6$ & 1.5588 & 1.5006 & 30.8 & 1.7 & - \\
CsCu(BiS$_2$)$_2$ & 0.2066 & 0.2056 & 81.2 & 75.5 & *\\
NaBiS$_2$ & 0.3454 & 0.3375 & 84.1 & 92.5 & - \\ 
Ga$_{12}$Ag$_2$Te$_{19}$ & 0.73 & 0.6685 & 0.0 & 30.8 & - \\
KI$_3$ $\cdot$ H$_2$O & 1.7388 & 1.693 & 22.6 & 8.7 & + \\
TlNO$_2$ & 1.9635 & 1.6023 & 132.4 & 9.8 & - \\
Cs$_2$Se & 2.0306 & 1.7223 & 0.0 & 41.1 & * \\
PbS (bilayer) & 1.2678 & 1.2678 & 36.0 & 18.5 & * \\
LaMoN$_3$ & 1.3212 & 1.1059 & 95.0 & 0.0 & - \\
Sb$_2$Pb$_2$O$_7$ & 1.6772 & 1.4991 & 29.5 & 0.0 & + \\
In$_2$Te$_5$ & 0.8637 & 0.7409 & 19.4 & 0.0 & - \\
La$_3$AgGeS$_7$ & 1.9512 & 1.9359 & 20.9 & 5.5 & + \\
Hf$_2$N$_2$O & 2.2349 & 2.0904 & 0.0 & 26.0 & - \\
Hg$_3$AsS$_4$Br & 1.6323 & 1.4992 & 0.0 & 38.7 & + \\
LaCrAgO$_6$ & 0.7335 & 0.7141 & 0.0 & 41.4 & -\\

 \hline
\end{tabular}
\caption{Selected noncentrosymmetric materials with large spin-orbit coupling, obtained from querying the \textsc{Materials Project} database. Tabulated are the direct and indirect band gaps, the spin-orbit splitting at the conduction band minimum ($\Delta E_{\text{CBM}}$) and at the valence band maximum ($\Delta E_{\text{VBM}}$), and the sign of the lifetime enhancement factor. ``$+$" indicates that the ordering of spin-split valence and conduction bands results in carrier lifetime enhancement, while ``$-$" indicates lifetime reduction, and ``*" indicates that the lifetime enhancement or reduction is dependent on the direction of the $k$-vector from the band minima. See Fig~\ref{fig:recomb} for the definition of valence and conduction band splittings.  }
\label{table:1}
\end{table}

\section{materials displaying spin-orbit modified carrier lifetime}

In Ref.~\cite{Zheng15p7794}, we showed that dynamic structural fluctuations play in important role in enhancing the carrier lifetime enhancement in MAPbI$_3$. In the high-symmetry structure of tetragonal phase MAPbI$_3$, the order of Rashba-split bands increases carrier recombination. However, structural fluctuations at room temperature change the ordering of bands to decrease carrier recombination instead. This is particularly important in the hybrid perovskites because of their weak bonds and dynamically fluctuating molecular components. For the materials presented below, we discuss their TDM properties in their most stable static structures, because dynamical fluctuation is not a necessary ingredient for a material to display spin-orbit-modified carrier lifetime.

\begin{figure}
\includegraphics[width=0.85\textwidth]{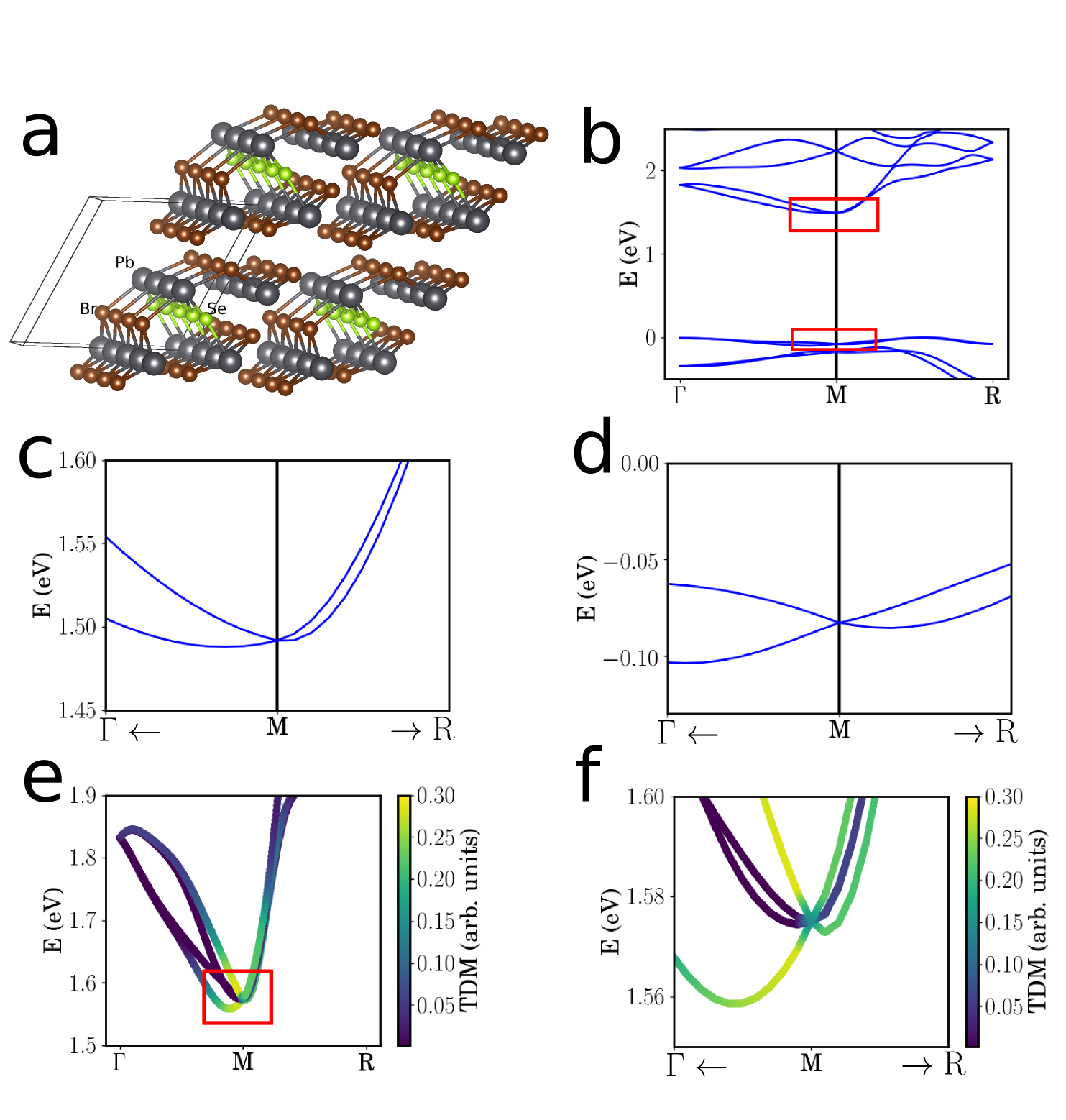}
  \caption{a) Crystal structure of Pb$_4$SeBr$_6$, with primitive unit cell depicted. Colored spheres represent Pb (grey), Se (light green), and Br (brown). b-d) DFT band structure of Pb$_4$SeBr$_6$, showing conduction (c) and valence (d) bands. e-f) Transition dipole moments of the four lowest-energy transitions between valence and conduction bands, with color representing the magnitude of transition dipole moments. f) Magnified view of transition dipole moments near the band edge.   }\label{fig:pb4sebr6}
\end{figure}

Pb$_4$SeBr$_6$ in its Imm2 structure consists of one-dimensional chains aligned along the $a$-axis (Fig.~\ref{fig:pb4sebr6}a). It has a Rashba band structure at the M point, with the largest splitting of the conduction band of 30.8 meV occurring along the $\Gamma$-M direction (Fig.~\ref{fig:pb4sebr6}c), and the largest splitting of the valence band of 1.7 meV occuring along the same direction. At the conduction band minimum, which is also located along $\Gamma$-M, we find that carrier recombination of the VB1$\rightarrow$CB1 transition is enhanced relative to the VB2$\rightarrow$CB1 and VB1$\rightarrow$CB2 transitions (Fig.~\ref{fig:pb4sebr6}e). At the CBM, the TDM of the VB1$\rightarrow$CB1 transition is 2530 times the TDM of the VB2$\rightarrow$CB1 transition. 

\begin{figure}
\includegraphics[width=0.85\textwidth]{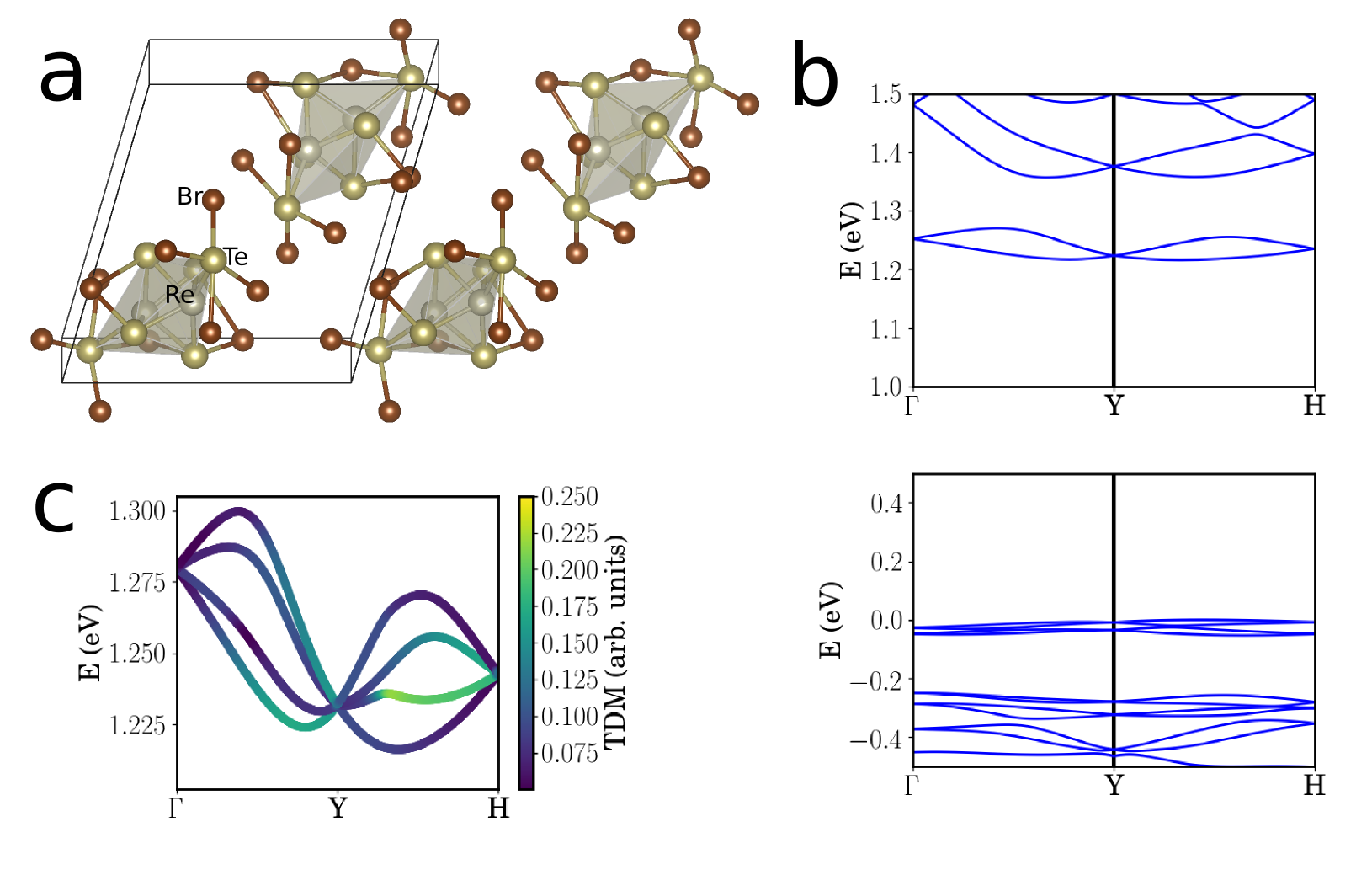}
  \caption{a) Crystal structure of ReTe$_3$Br$_5$, with primitive unit cell depicted. Colored spheres represent Re (silver), Te (gold), and Br (brown). b) DFT band structure of ReTe$_3$Br$_5$, showing conduction (top) and valence (bottom) bands.  c) Transition dipole moments of the four lowest-energy transitions between valence and conduction bands, with color representing the magnitude of transition dipole moments.   }\label{fig:rete3br5}
\end{figure}

ReTe$_3$Br$_5$ has the opposite ordering of bands at its band edges. The P2$_1$ phase of this material is made up of free-standing Re$_2$Te$_6$Br$_{10}$ octahedral units, with Te located at the vertices of the octahedra and the two Re located at their interior (Fig.~\ref{fig:rete3br5}a). Each apex of the Te$_6$ octahedron is bonded to four Br, while the remaining two Br are bonded to Te at the equatorial plane of the octahedron.  The conduction band has a spin-orbit splitting of 27.6 meV (Fig.~\ref{fig:rete3br5}b), while the valence band has a splitting of 6.3 meV.  Fig.~\ref{fig:rete3br5}c shows that recombination of the VB1$\rightarrow$CB1 transition is suppressed at the band edges, which are located along the Y-H line. In this material, the ratio of the TDMs of the lowest-energy VB1$\rightarrow$CB1 transition to the higher-energy  VB2$\rightarrow$CB1 transition is 0.426.

Besides affecting the TDMs, a secondary effect of spin-orbit coupling is to shift the locations of the VBM and CBM. In Pb$_4$SeBr$_6$, the CBM is only shifted slightly away from the M point (Fig.~\ref{fig:pb4sebr6}c), whereas it is shifted to a much greater extent along the Y-R line in ReTe$_3$Br$_5$, where the spin-orbit splitting is so large that it is comparable to the band width. The VBM and CBM will generally be shifted to different positions, resulting in an indirect band gap. This is another factor which suppresses the recombination rate. The extent of this suppression depends on the energy scale over which the band gap is indirect. If the difference between the direct and indirect band gaps is not much bigger than the room temperature energy scale (such as in Pb$_4$SeBr$_6$ and ReTe$_3$Br$_5$), the indirect gap effect will be masked by the Fermi distribution of carriers. Some materials, such as Cs$_2$Se (Table.~\ref{table:1}), have a much greater difference between direct and indirect gaps, and will display this effect to a greater extent. We note that in these spin-orbit materials, it is rare for the difference between direct and indirect gaps to be so large that it affects absorption of visible light. They thus act effectively as direct gap semiconductors when absorbing light, and as indirect gap semiconductors for carrier recombination.

\begin{figure}
\includegraphics[width=0.85\textwidth]{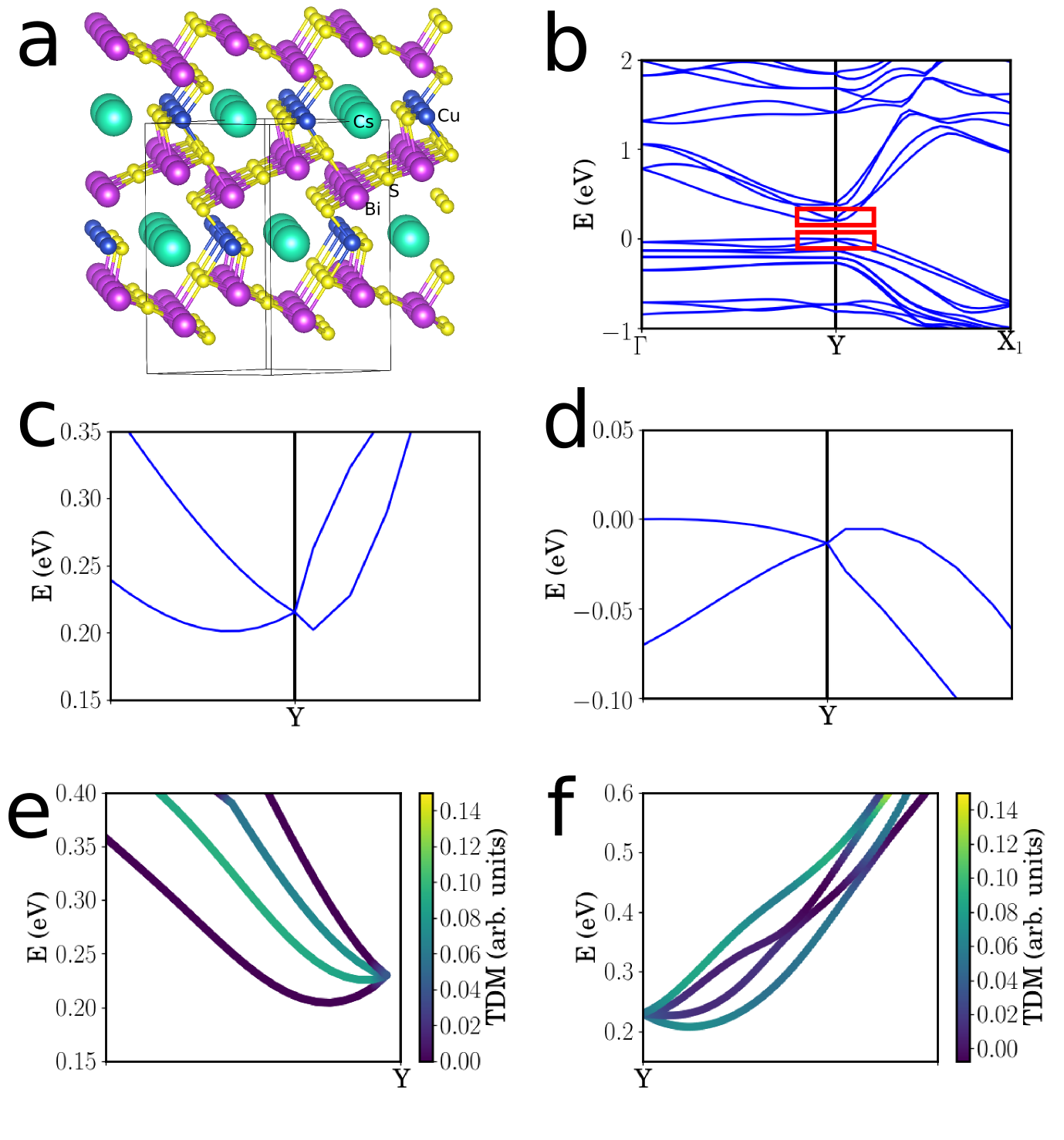}
  \caption{a) Crystal structure of CsCu(BiS$_2$)$_2$, with primitive unit cell depicted. Colored spheres represent Cs (cyan), Cu (blue), Bi (magenta), and S (yellow). b-d) DFT band structure of CsCu(BiS$_2$)$_2$, showing conduction (c) and valence (d) bands. e-f) Transition dipole moments of the four lowest-energy transitions between valence and conduction bands, with color representing the magnitude of transition dipole moments, along the $\Gamma$-Y direction (e), and along the Y-X$_1$ direction (f).  }\label{fig:cscubis}
\end{figure}

Compared to Pb$_4$SeBr$_6$ and ReTe$_3$Br$_5$, the band structure and TDM structure of CsCu(BiS$_2$)$_2$ is more complex. This material, which belongs to space group Cmc2$_1$, is made up of a Cu(BiS$_2$)$_2$ network which has one-dimensional channels, occupied by Cs atoms (Fig.~\ref{fig:cscubis}a). There is a direct band gap at the Y point. Both conduction and valence bands display large spin-orbit splittings, of 81.2 meV and 75.5 meV respectively. We find that carrier recombination is suppressed along the $\Gamma$-Y direction (by a factor of 0.00102), but is enhanced along the Y-X$_1$ direction (by a factor of 3.29). This example shows that recombination enhancement or reduction can be a function of $k$-vector direction. In this situation, we expect carriers to accumulate in the sections of the band edges where recombination is reduced, here, along the $\Gamma$-Y direction. This will lead to an anisotropic $k$-space distribution of carriers, which can be detected by time-resolved excited-state optical measurements. 

\section{Conclusions}

In conclusion, we have shown that materials with large spin-orbit coupling and broken inversion symmetry can have their TDMs, and hence carrier recombination rates, strongly modified by spin-orbit coupling. Our DFT calculations have shown that carrier recombination can either be enhanced, reduced, or redistributed in $k$-space by this mechanism. While the halide perovskites are thus far the only class of materials displaying experimental evidence for this effect, we have shown that there are potentially many more materials that would as well. This raises the possibility for the targeted use of spin-orbit materials in optoelectronics applications and for observing excited-state photophysics. 

\section{Acknowledgements}
L.Z.T. was supported by the U.S. ONR under Grant N00014-17-1-2574.
A.M.R. was supported by the U.S. Department of Energy, under grant DE-FG02-07ER46431.
Computational support was provided by the HPCMO of the U.S. DOD and the NERSC of the U.S. DOE.




\clearpage

\bibliography{rappecites,newcites}

\end{document}